\documentclass[aps,pra,superscriptaddress,floatfix,twocolumn,showpacs]{revtex4-1}
\usepackage{amsmath}
\usepackage{amsfonts}
\usepackage{amssymb}
\usepackage{graphicx}
\usepackage{bm}
\usepackage{color}
\usepackage[hypertex]{hyperref}
\usepackage[active]{srcltx}
\usepackage{cases}
\usepackage{ulem}
\usepackage{multirow}

\newcommand{\be}{\begin{equation}}
\newcommand{\ee}{\end{equation}}
\newcommand{\bea}{\begin{eqnarray}}
\newcommand{\eea}{\end{eqnarray}}

\newcommand{\PT}{$\mathcal{PT}$ }
\newcommand{\ket}[1]{|{#1}\rangle}
\newcommand{\bra}[1]{\langle{#1}|}
\newcommand{\mP}{\mathcal{P}}
\newcommand{\mT}{\mathcal{T}}
\newcommand{\mK}{\mathcal{K}}
\newcommand{\mPT}{\mathcal{PT}}

\usepackage{graphicx}
\usepackage{bm}

\newcommand{\ColorOnline}{(Color online) }

\def\stacksymbols #1#2#3#4{\def\theguybelow{#2}
    \def\verticalposition{\lower#3pt}
    \def\spacingwithinsymbol{\baselineskip0pt\lineskip#4pt}
    \mathrel{\mathpalette\intermediary#1}}
\def\intermediary#1#2{\verticalposition\vbox{\spacingwithinsymbol
      \everycr={}\tabskip0pt
      \halign{$\mathsurround0pt#1\hfil##\hfil$\crcr#2\crcr
               \theguybelow\crcr}}}

\begin{document}
\title{
Explicit definition of \PT symmetry for non-unitary quantum walks with gain and loss
}

\author{Ken Mochizuki}
\affiliation{
Department of Applied Physics, Hokkaido University, Sapporo 060-8628, Japan
}

\author{Dakyeong Kim}
\affiliation{
Department of Applied Physics, Hokkaido University, Sapporo 060-8628, Japan
}

\author{Hideaki Obuse}
\affiliation{
Department of Applied Physics, Hokkaido University, Sapporo 060-8628, Japan
}

\date{May 26, 2016}

\begin{abstract}
\PT symmetry, that is, a combined parity and time-reversal symmetry
 is a key milestone for non-Hermite systems exhibiting entirely real
 eigenenergy. In the present work, motivated by a recent experiment, we
 study \PT symmetry of the time-evolution operator of non-unitary
 quantum walks. We present the explicit definition of \PT symmetry by
 employing a concept of symmetry time frames. We provide a necessary and
 sufficient condition so that the time-evolution operator of the
 non-unitary quantum walk retains \PT symmetry even when parameters of the model depend on position. It is also shown that there exist extra symmetries embedded in the time-evolution operator. Applying these results, we clarify that the non-unitary quantum walk in the experiment does have \PT symmetry.
\end{abstract}


\maketitle
\section{introduction}
Quantum mechanics requires that, in a closed system, physical observables be represented by Hermitian operators. The Hamiltonian of the system is no exception to this rule. However, the closed system is an ideal concept and, rigorously speaking, almost systems in the real world, except a whole universe, should have flow of energy and particles to outer environments, which makes the Hamiltonian of the inner system non-Hermitian. Furthermore, it is widely accepted to phenomenologically include non-Hermite effects into Hamiltonians when we treat effects of amplification and dissipation, namely, gain and loss,
in open systems. For example, non-Hermitian Hamiltonians are employed to
describe the radioactive decay\cite{nh1}, the depinning of flux lines in type-II superconductors\cite{nh2}, and so on\cite{nh6}. In general, such a non-Hermitian Hamiltonian has complex eigenenergy which makes systematic controls of the dynamics difficult.

In 1998, however, Bender and Boettcher clarified that a broad class of non-Hermitian Hamiltonians can have entirely real eigenenergy if the
system possesses a combined parity symmetry and time-reversal symmetry
(TRS), that is, \PT symmetry\cite{pt1,Bender99,pt2,pt3}. If the Hamiltonian possesses \PT symmetry and its eigenstates are also eigenstates of the $\mathcal{PT}$ symmetry operator, then, this is a sufficient condition for the eigenenergy being real. Applying this property, moreover, \PT symmetry in
the non-Hermitian Hamiltonian provides a procedure to selectively induce
complex eigenenergy for specific eigenstates\cite{pt4,pt5,pt6}. For systems
described by non-Hermitian Hamiltonians with \PT symmetry, a large
number of novel phenomena, which can not be observed in Hermitian
systems, have been predicted theoretically. For example, the system with
\PT symmetric periodic structures can act as unidirectional invisible
media\cite{pt7,pt8}, edge states having complex eigenenergy emerge\cite{pt9,pt10}, Bloch oscillations with unique features occur\cite{pt11}, and others\cite{pt12,pt13,pt14,pt15,pt16,pt17,pt18,pt19,Longhi10}. These results open a way to engineer non-Hermite systems to utilize as novel platforms of applications. The system with \PT symmetry has been realized in optics by using coupled optical waveguides with fine tuned complex refractive index\cite{pt20,pt21}. It has been also demonstrated that coupled microcavity resonators realize \PT symmetric systems\cite{pt22,pt23}. Recently, the mode-selective lasing by utilizing \PT symmetry has been realized based on microring resonators\cite{pt24,pt25}. However, due to difficulty to handle gain and loss effects, the experimental systems are limited to a small number of elements.

In contrast, there is a unique way to experimentally perform large scale
\PT symmetric systems with high tunability, that is, the discrete-time
quantum walk\cite{kempe03,ambainis03}. The discrete-time quantum walk
(quantum walk, in short) is the model recently attracting attention as
versatile platforms for quantum computations and quantum simulators. The quantum walk describes quantum dynamics of particles by a time-evolution operator, instead of a Hamiltonian. The quantum walks have been realized in various experimental setups, such as cold atoms\cite{qw1}, trapped ions\cite{qw2,qw3}, and optical systems\cite{qw4,qw5,regb4,qw6,qw7}. Since quantum walks enable high tunability of the system setup, various phenomena which require delicate setups have been observed, such as Anderson localization\cite{qw8,qw9}, scattering with positive- and negative-mass pulses\cite{regb5}, emergence of edge states which stem from topological phases\cite{qw12}, and so on.

Remarkably, in 2012, a quantum walk by optical-fibre loops, where additional optical amplifiers make it possible to control the effects of gain and
loss was experimentally implemented\cite{regb1}. Due to gain and loss, the time-evolution operator of this quantum walk becomes non-unitary, which can be
considered that the effective Hamiltonian is non-Hermitian. Nevertheless, it has been shown that the system has entirely real (quasi-)energy in proper setups. Furthermore, interesting phenomena peculiar to \PT symmetry, such as unidirectional invisible transport\cite{regb1}, extraordinary Bloch oscillations\cite{regb1}, optical solitons\cite{regb2,regb3}, have been
observed. These results provide convincing evidence that the system
possesses \PT symmetry. However, \PT symmetry and the \PT symmetry operator have not yet been directly identified from the time-evolution operator itself, since the definition of \PT symmetry on the time-evolution operator has not been established so far. It is an urgent and important task to identify the explicit definition of \PT symmetry for further developments.

In the present work, we provide the explicit definition
of the \PT symmetry operator and identify that the time
evolution operator of the non-unitary quantum walk in the
experiment has, indeed, \PT symmetry. This is archived for the first
time by employing a concept of symmetry time frames\cite{qw14} which has been developed in the recent study of topological phases of quantum walks\cite{qw14,qw10,qw11,qw13,qw15}. We also show that the time-evolution operator of the non-unitary quantum walk has extra symmetries. Furthermore, we provide the necessary and sufficient conditions for \PT and other
symmetries of the time-evolution operator even when parameters of the model are
position dependent. Taking account of these results, we present inhomogeneous
non-unitary quantum walk with \PT symmetry. (We note that, although
the argument on \PT symmetry to retain reality of (quasi-) energy has been generalized in Refs.\ \cite{bender02,mostafazadeh02,mostafazadeh04}, we focus  on \PT symmetry in the original sense of Ref.\ \cite{pt1} in the present work.)

This paper is organized as follows. We define the time-evolution operator of the non-unitary quantum walk in Sec. \ref{sec:model}. Section \ref{sec:paes} is devoted to present how to define and identify \PT symmetry and extra symmetries of the time-evolution operator of the non-unitary quantum walk.  
This is our main result of the present work. In Sec.\
\ref{sec:applications}, as applications of the result obtained in the previous
section, we identify \PT symmetry of the time-evolution operator of the
non-unitary quantum walk in the experiment\cite{regb1} and, further,
demonstrate a \PT symmetric inhomogeneous non-unitary quantum walk.
The summary and discussions are given in Sec.\ \ref{sec:summary}.

\section{Definition of time-evolution operators of non-unitary quantum walks}
\label{sec:model}

\begin{figure}[t]
\begin{center}
\includegraphics[scale=0.30]{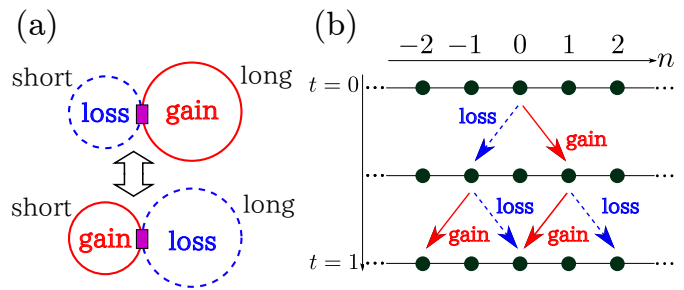}
\caption{\ColorOnline  (a)Experimental setup. Optical pulses
 corresponding to walkers go around in two optical-fibre loops with
 different circumferences, and they are split into two at the connected
 point (shown by a rectangle) corresponding to coin operators. After a
 single cycle, pulses are delayed or advanced in time due to the
 difference of lengths of two fibre loops, corresponding to shift operators. The time evolution of the single time step is composed of the following two substeps. At the former half of the step, amplitudes of pulses passing through the long (short) loop are amplified (dumped) and, at the latter half of the step, vice verse. (b)Translation from the above description to the standard schematic view of the 1D two-step quantum walk. When a pulse passes the long (short) loop and it is delayed (advanced) in time, this is interpreted as that the walker "shifts to right (left)". In both (a) and (b), loops or arrows with gain (loss) are depicted in solid (dashed) lines.}
\label{fig:es}
\end{center}
\end{figure}

Figure \ref{fig:es} shows the schematic view of the experimental setup of
the non-unitary quantum walk implemented by the two optical-fibre loops
in Ref. \cite{regb1}. As explained in the caption, the system is interpreted as
one-dimensional (1D) two-step quantum walks. Motivated by the
experiment, we define a time-evolution operator of the non-unitary
quantum walk with gain and loss so that one can flexibly tune various
parameters of the system, while the basic setup of the system
should not be altered. At first, we introduce the time-evolution operator of the 1D two-step unitary quantum walk, and then extend it to the non-unitary one. We introduce the basis of the walker's 1D position space $\ket{n}$ and internal states $\ket{L}=(1,0)^{T}, \ket{R}=(0,1)^{T}$ where the subscript $T$ denotes the transpose. The symbols $L,R$ represent walker's internal states, say, left mover and right mover components, respectively. The time-evolution operator of the two-step unitary quantum walk $U_{u}$ is defined as
\begin{align*}
U_{u}=S\,C(\theta_{2})\,S\,C(\theta_{1}).
\end{align*}
Here, the coin operator $C(\theta_i)$, where the
subscript $i=1$ or $2$ distinguishes the parameter for the first or second coin operators, respectively, and the shift operator $S$ are standard elemental operators of quantum walks defined as
\begin{subequations}
\begin{align}
&C(\theta_{i})=\sum_{n}\ket{n}\bra{n}\otimes\tilde{C}(\theta_{i,n}),
\label{chx}\\
&\tilde{C}(\theta_{i,n})=\left(
 \begin{array}{cc}
 \cos[\theta_{i}(n)]&i\sin[\theta_{i}(n)] \\
 i\sin[\theta_{i}(n)]&\cos[\theta_{i}(n)]
 \end{array} 
\right),
\label{ctx}
\end{align}
\end{subequations}
and
\begin{align}
S=\sum_{n}\left(
 \begin{array}{cc}
 \ket{n-1}\bra{n} &0 \\
 0&\ket{n+1}\bra{n} 
 \end{array} 
\right).
\label{shx}
\end{align}
Since $\tilde{C}(\theta_{i,n})$ acts on the internal states of walkers at the position $n$, the coin operator $C(\theta_{i})$ mixes the walker's internal states, where the value of $\theta_{i}(n)$ determines how strongly to mix at each position $n$. The shift operator $S$ changes the position of walkers depending on the internal states. Note that, in the present work, we follow a rule that an operator with a tilde $(\tilde{\,\,})$ on the top acts on space of internal states of walkers.

With an initial state $\ket{\psi(0)}$, the wave function after $t$ time step is described as
\begin{align*}
\ket{\psi(t)}=U^{t}\ket{\psi(0)}=\sum_{n,\sigma=L,R}\psi_{n,\sigma}(t)\,\ket{n}\otimes\ket{\sigma}.
\end{align*}
From the eigenvalue equation, we define the quasi-energy $\varepsilon$ as
\begin{align*}
U\ket{\Psi_{\lambda}}=\lambda\ket{\Psi_{\lambda}},\ 
\lambda=e^{-i\varepsilon},
\end{align*}
where $\ket{\Psi_{\lambda}}$ is the eigenvector with the eigenvalue $\lambda$. For the unitary quantum walk, $\lambda$ should satisfy $|\lambda|=1$ and then $\varepsilon$ should be real with $2\pi$ periodicity. 
 
\begin{figure}[t]
\begin{center}
\includegraphics[scale=0.30]{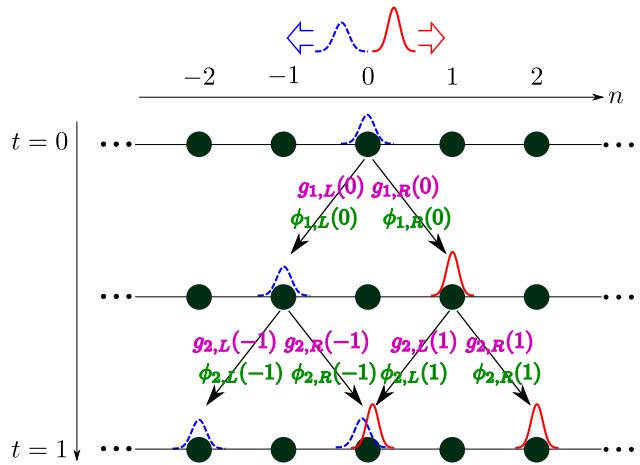}
\caption{\ColorOnline An example of the one time step of time evolutions
 described by the time-evolution operator $U$ with the initial state $\ket{\psi(0)}=\ket{0,L}$. The left (right) mover components are depicted as waves in dashed (solid) curves. At each time step, on a site $n$, the left (right) mover component $\psi_{n,L(R)}(t)$  is varied to the linear combination of $\psi_{n,L}(t)$ and $\psi_{n,R}(t)$ by the coin operator $C(\theta_{i})$.  Then, left mover components $\psi_{n,L}(t)$ move to left and right mover components $\psi_{n,R}(t)$ move to right by the shift operator $S$. During walkers change their positions, they are affected by gain or loss of the amplitude and phase modulation, that is, $\psi_{n,\sigma}(t)$ increase or decrease by the factor $g_{i,\sigma}(n)$  by the gain-loss operator $G_{i}$, and earn the phase $\phi_{i,\sigma}(n)$ by the phase operator $\Phi_{i}$.}
\label{fig:te}
\end{center}
\end{figure}

The unitary quantum walk described by $U_{u}$ can be extended to the non-unitary one described by
\begin{align}
U=S\,G_{2}\,\Phi_{2}\,C(\theta_{2})\,
S\,G_{1}\,\Phi_{1}\,C(\theta_{1}),
\label{uhx}
\end{align}
which is consistent with the basic experimental setup in Ref. \cite{regb1}. Here, we introduce additional elemental operators; the gain-loss operator $G_{i}$ and the phase operator $\Phi_{i}$ defined as
\begin{align}
&G_{i}=\sum_{n}\ket{n}\bra{n}\otimes\tilde{G}_{i,n},\ 
\tilde{G}_{i,n}=\left(
 \begin{array}{cc}
 g_{i,L}(n)&0 \\
 0&g_{i,R}(n)
 \end{array} 
\right),
\label{gx}\\
&\Phi_{i}=\sum_{n}\ket{n}\bra{n}\otimes\tilde{\Phi}_{i,n},\ 
\tilde{\Phi}_{i,n}=\left(
 \begin{array}{cc}
 e^{i\phi_{i,L}(n)}&0 \\
 0&e^{i\phi_{i,R}(n)}
 \end{array}
\right),
\label{phix}
\end{align}
respectively. The gain-loss operator $G_{i}$ multiplies the wave function amplitude $\psi_{n,\sigma}(t)$ by the factor $g_{i,\sigma}(n)$. If $g_{i,\sigma}(n)\neq1$, then $G_{i}$ and $U$ become non-unitary operators. The phase operator $\Phi_{i}$ adds the phase $\phi_{i,\sigma}(n)$ to that of the wave function amplitude $\psi_{n,\sigma}(t)$. The time-evolution of a walker described by $U$ is schematically explained in Fig \ref{fig:te}.Thereby, the time-evolution operator of the non-unitary quantum walk contains three kinds of $n$ dependent parameters, $\theta_{i}(n)$, $g_{i,\sigma}(n)$, and $\phi_{i,\sigma}(n)$. The setup in the experiment in Ref.\ \cite{regb1} is realized with the parameters in Eq.\ (\ref{pex}), as we discuss in Sec.\ \ref{sec:applications}.

\section{\PT and extra symmetries of the non-unitary quantum walk}
\label{sec:paes}
In this section, we identify various symmetries embedded in the
time-evolution operator of the non-unitary quantum walk in Eq.\
(\ref{uhx}). Among them, our main target is \PT symmetry, which can restrict the quasi-energy of the non-unitary quantum walk to real number. To begin with, let us summarize argument on \PT symmetry of Hamiltonians\cite{pt1}. In
order to define \PT symmetry, we consider parity symmetry and TRS at first. For a system described by a (Hermitian or non-Hermitian) Hamiltonian $H$, it is required that the Hamiltonian satisfies following relations to retain parity symmetry and TRS:
\begin{align}
&\mP H\mP^{-1}=H,
\label{ph}\\
&\mT H\mT^{-1}=H,
\label{th}
\end{align}
respectively. Here, the parity symmetry operator $\mP$, which flips the sign of
position from $n$ to $-n$, is a unitary operator and does not include complex conjugation $\mK$. The TRS operator $\mT$, which inverts the direction of time from $t$ to $-t$, is an anti-unitary operator including $\mK$. By combing Eqs. (\ref{ph}) and (\ref{th}), \PT symmetry of the Hamiltonian is defined as
\begin{align}
(\mPT)H(\mPT)^{-1}=H,
\label{pth}
\end{align}
where the combined symmetry operator \PT is the anti-unitary operator.

When the Hamiltonian satisfies both Eqs.\ (\ref{ph}) and (\ref{th}), the
relation for \PT symmetry (\ref{pth}) is automatically satisfied. However, even when the Hamiltonian have neither parity symmetry [Eq.\ (\ref{ph})] nor TRS [Eq.\ (\ref{th})], it could satisfy Eq.\ (\ref{pth}) to establish \PT symmetry. This recovering of \PT symmetry becomes much important in the case of non-Hermitian Hamiltonians, since one of the standard ways to phenomenologically include effects of gain and loss is adding non-Hermite imaginary potential terms into a Hermitian Hamiltonian, which prevents to retain TRS in Eq.\ (\ref{th}) due to complex conjugation $\mK$. In addition to the presence of \PT symmetry of the non-Hermitian Hamiltonian, we demand that eigenvectors of the non-Hermitian Hamiltonian are also eigenvectors of the \PT symmetry operator,
\begin{align}
H \ket{\Psi_\lambda} = E_\lambda \ket{\Psi_\lambda},\quad\,
\mPT \ket{\Psi_\lambda} = e^{i \delta} \ket{\Psi_\lambda},
\label{eq:pt eigenvector}
\end{align} 
where the phase $\delta$ is a real number. 
Satisfying both conditions Eqs.\ (\ref{pth}) and (\ref{eq:pt eigenvector})
establishes the sufficient condition that the eigenenergy $E_{\lambda}$ is kept to be a real number even for the non-Hermitian Hamiltonian. Hereafter, we apply the above argument to the time-evolution operator of non-unitary quantum walks.

\subsection{Symmetries in homogeneous systems}
\label{sec:sihs}
For simplicity, at first, we assume the homogeneous non-unitary quantum walk in which all parameters have no position $n$ dependences, so that we can treat operators in momentum space by applying the Fourier transformation.

In the homogeneous systems, the operators, $C(\theta_{i})$, $G_{i}$, and $\Phi_{i}$, are diagonal in the momentum representation, and we can drop the subscript $n$ from $\tilde{C}(\theta_{i,n})$, $\tilde{G}_{i,n}$, and $\tilde{\Phi}_{i,n}$. For further simplification, we assume
\begin{align}
&\tilde{G}_{2}=\tilde{G}^{-1}_{1}=\tilde{G}=\left(
 \begin{array}{cc}
 e^{\gamma}&0 \\
 0&e^{-\gamma}
 \end{array} 
\right)=e^{\gamma\sigma_{3}},
\label{gtk}\\
&\tilde{\Phi}_{2}=\tilde{\Phi}_{1}=\tilde{\Phi}=\left(
 \begin{array}{cc}
 e^{i\phi}&0 \\
 0&e^{-i\phi}
 \end{array} 
\right)=e^{i\phi\sigma_{3}},
\label{phitk}
\end{align}
where $\sigma_{j=1,2,3}$ are Pauli matrices. [The peculiar choice of
$\tilde{G}_2 = \tilde{G}_1^{-1}$ is motivated by the setup of the
experiment\cite{regb1} as shown in Eqs.\ (\ref{gle}) and (\ref{gre}).] By using the Pauli matrix $\sigma_{1}$, the coin operator is also written as
\begin{align}
\tilde{C}(\theta_{i})=\left(
 \begin{array}{rr}
 \cos[\theta_{i}]&i\sin[\theta_{i}] \\
 i\sin[\theta_{i}]&\cos[\theta_{i}]
 \end{array} 
\right)=e^{i\theta_{i}\sigma_{1}}.
\label{ctk}
\end{align}
With the Fourier transformation, the shift operator in Eq. (\ref{shx}) can be rewritten as
\begin{align}
S=\sum_{k}\ket{k}\bra{k}\otimes\tilde{S}(k),\ 
\tilde{S}(k)=\left(
 \begin{array}{cc}
e^{+ik}&0 \\
 0&e^{-ik}
 \end{array}
\right)=e^{ik\sigma_{3}},
\label{sk}
\end{align}
where $k$ stands for the wave number. Accordingly, the time-evolution
operator $U$ in Eq. (\ref{uhx}) in the momentum representation is
written down as
\begin{subequations}
\begin{align}
&U=\sum_{k}\ket{k}\bra{k}\otimes\tilde{U}(k),
\label{uhk}\\
&\tilde{U}(k)=\tilde{S}(k)\,\tilde{G}\,\tilde{\Phi}\,\tilde{C}(\theta_{2})\,\tilde{S}(k)\,\tilde{G}^{-1}\,\tilde{\Phi}\,\tilde{C}(\theta_{1}).
\label{utk}
\end{align}
\label{uk}\end{subequations}
Since determinants of all the above elemental operators are one, the
determinant of the time-evolution operator $\tilde{U}(k)$ is also one,
while the operator is non-unitary when $\gamma \ne 0$.

\begin{figure}[tbh]
\begin{center}
\includegraphics[width=7.1cm]{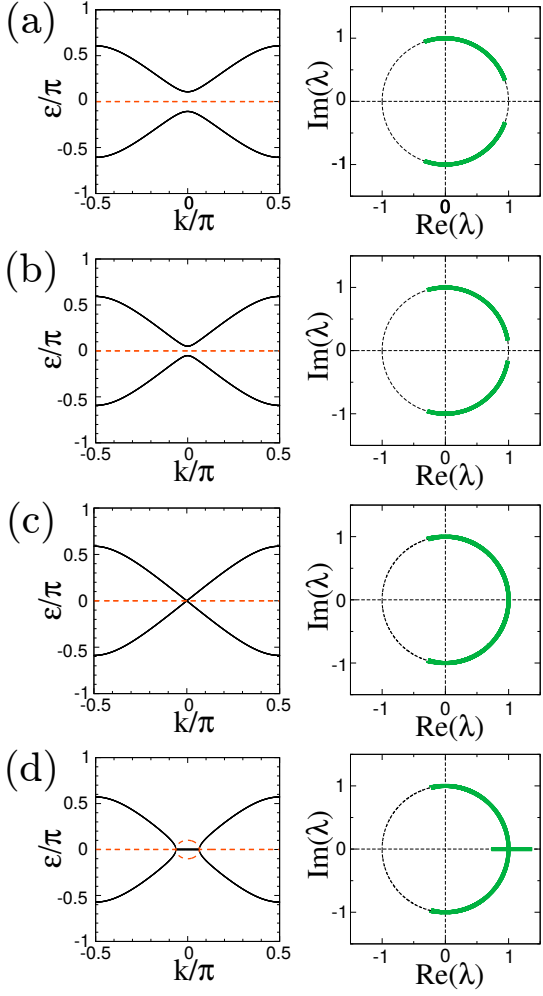}
\vspace{-0.2cm}
\caption{\ColorOnline The quasi-energy in Eq.\ (\ref{qe}) with various gain-loss parameters when $\theta_{1}=\pi/4,\ \theta_{2}=-\pi/7$, and $\phi=0$. The left column shows the quasi-energy as a function of $k$ where the solid (dashed) curves represent the real (imaginary) part of the quasi-energy, while
 the  right column shows the eigenvalue on a unit circle indicating $|\lambda|=1$ on a complex plain. (a)In the case of $e^{\gamma}=1$, all of the quasi-energy are real as the time-evolution operator is unitary. (b)In the case of $e^{\gamma}=1.3$, the quasi-energy is entirely real although the time-evolution operator is non-unitary, and quasi-energy gaps around $\varepsilon=0,\pi$ open. (c)In the case of $e^\gamma=\exp\{\cosh^{-1}[(\cos\theta_{1}\cos\theta_{2}-1)/(\sin\theta_{1}\sin\theta_{2})]/2\}=1.34\cdots$, while quasi-energy is entirely real, the quasi-energy gap around $\varepsilon=0$ closes. (d)In the case of $e^{\gamma}=1.5$, the quasi-energy becomes complex for $|k|/\pi \lesssim 0.1$, and the gap closes.}
\label{fig:qe}
\end{center}
\end{figure}

By solving the eigenvalue problem, the quasi-energy of the time-evolution operator in Eq. (\ref{utk}) is derived as
\begin{align}
\cos(\pm \varepsilon)=\cos\theta_{1}\cos\theta_{2}\cos2(k+\phi)
-\sin\theta_{1}\sin\theta_{2}\cosh(2\gamma),
\label{qe}
\end{align}
and the corresponding eigenvector is
\begin{align}
\ket{\Psi_{k,\pm}}&=
 e^{-i \frac{\theta_1}{2} \sigma_1}
\frac{  e^{-i \eta_k}
}{2\sqrt{\cos2\xi_{k}}}
\left(\begin{array}{c}
e^{i\alpha} \pm e^{- i\alpha}\\
-i\left[
e^{i\alpha} \mp e^{-i\alpha}
\right]
\end{array}\right),
\label{eq:eigenvector}\\
\alpha&=\eta_k \pm \xi_k,\nonumber
\end{align}
where $\eta_k$ and $\xi_k$ are defined as
\begin{align*}
\tan(2\eta_k)&=d_1/d_3,\nonumber \\
\cos(2\xi_{k})&=\sqrt{1-(d_{2}/|d_{k}|)^{2}},\quad \sin(2\xi_{k})=d_{2}/|d_{k}|,\nonumber\\
|d_{k}|&=|d_{3}+id_{1}|,\nonumber\\
d_{1}&=\sin\theta_{1}\cos\theta_{2}\cos2(k+\phi)+\cos\theta_{1}\sin\theta_{2}\cosh(2\gamma),\nonumber\\
d_{2}&=-\sin\theta_{2}\sinh(\pm2\gamma),\\
d_{3}&=-\cos\theta_{2}\sin2(k+\phi).\nonumber
\end{align*}
We remark that, while $\eta_k$ is always real, $\xi_k$ becomes imaginary
when $d_2^2 > d_1^2+d_3^2$. Figure \ref{fig:qe} shows the quasi-energy
as a function of $k$ with several values of $\gamma$;
(a) $e^\gamma=1$ , (b) $e^\gamma=1.1$, (c) $e^{\gamma}=1.34\cdots$, and (d) $e^\gamma=1.5$ (see the caption of Fig.\ \ref{fig:qe} for other parameters).
Comparing with the case of the unitary quantum walk in Fig.\ \ref{fig:qe}
(a), we see from Fig.\ \ref{fig:qe} (b) that , while the quasi-energy gap around $\varepsilon=0$ becomes narrow, the quasi-energy remains entirely real even for the finite $\gamma$ (non-unitary quantum walks). This keeps holding as long as the absolute value of the right hand side in Eq. (\ref{qe}) does not exceed one, which is consistent with the condition to keep $\xi_k$ real. The value of $\gamma$ used for Fig.\ \ref{fig:qe} (c) corresponds to this limit and the quasi-energy gap closes at $\varepsilon=0$, so-called the exceptional point\cite{Bender99}. When $\gamma$ exceeds this value, part of quasi-energy whose components of real number is zero exhibits finite values of imaginary number, as shown in Fig.\ \ref{fig:qe} (d). These observations suggest the presence of \PT symmetry or more generalized symmetries in Refs.\ \cite{bender02,mostafazadeh02,mostafazadeh04}. Henceforth, we
show that there exists \PT symmetry, as Ref. \cite{regb1} has stated. In addition, from Eq. (\ref{qe}), we also understand that the quasi-energy becomes symmetric with respect to $\varepsilon=0$. Indeed, these properties can be understood from symmetries embedded in the non-unitary time-evolution operator in Eq.\ (\ref{uk}), which is also shown in the following subsections. 

\subsubsection{\PT symmetry}
We introduce the parity symmetry and TRS operators, $\mP$
and $\mT$, in the position and momentum representations as follows;
\begin{align}
\mP & = \sum_n \ket{-n}\bra{n}\otimes \tilde{\mP} = \sum_k
 \ket{-k}\bra{k}\otimes \tilde{\mP},
\label{eq:parity}\\
\mT & = \sum_n \ket{n}\bra{n}\otimes \tilde{\mT} = \sum_k
 \ket{-k}\bra{k}\otimes \tilde{\mT},
\label{eq:TRS}
\end{align}
where $\tilde{\mP}$ and $\tilde{\mT}$ act on internal space of the
time-evolution operator.
We understand that the parity symmetry operator $\mP$ flips the sign of
momentum $k$ because the operator $\mP$ changes the position
$n$ to $-n$ and the TRS operator $\mT$ also flips the sign of $k$ since
the operator $\mT$ is an anti-unitary operator including a complex
conjugation $\mK$.

Then, we convert Eqs. (\ref{ph})-(\ref{pth}) for the Hamiltonian into those
for the time-evolution operator in Eq.\ (\ref{uk}). By using the relation between the time-evolution operator and the effective Hamiltonian: $U=e^{-iH}$, 
We derive relations for parity-symmetry, TRS, and \PT symmetry as
\begin{align*}
\mP U \mP^{-1}&=U,
\\
\mT U \mT^{-1}&=U^{-1},
\\
(\mP\mT)U(\mP\mT)^{-1}&=U^{-1}.
\end{align*}
By substituting Eqs.\ (\ref{eq:parity}) and (\ref{eq:TRS}) into the
above relations, we obtain 
\begin{align}
\tilde{\mP}\tilde{U}(k)\tilde{\mP}^{-1}&=\tilde{U}(-k),
\label{pu}\\
\tilde{\mT}\tilde{U}(k)\tilde{\mT}^{-1}&=\tilde{U}^{-1}(-k),
\label{tu}\\
(\tilde{\mP}\tilde{\mT})\tilde{U}(k)(\tilde{\mP}\tilde{\mT})^{-1}&=\tilde{U}^{-1}(+k),
\label{ptu}
\end{align}
respectively.

In order to identify symmetries, we need to examine whether the
time-evolution operator of the non-unitary quantum walk in Eq.\
(\ref{utk}) satisfies the above relations. 
For parity symmetry in Eq.\ (\ref{pu}), on one hand, we can straightforwardly
obtain relations for the same elemental operators by comparing left and right
hand sides of Eq.\ (\ref{pu}) by substituting Eq.\ (\ref{utk}), {e.g.}, $\tilde{\mP}
\tilde{S}(k) \tilde{\mP}^{-1} = \tilde{S}(-k)$, $\tilde{\mP}
\tilde{G} \tilde{\mP}^{-1} = \tilde{G}$, and etc. 
On the other hand, for TRS and \PT symmetry, 
there appear the inverse operators of the time-evolution operator in the
right hand side of Eqs.\ (\ref{tu}) and (\ref{ptu}), which invert the
time order of elemental operators and then prevent us from deriving the one to one correspondence for the same elemental operators. Indeed, according to recent work on symmetries which are important to topological phases of quantum walks, it has become clear that the presence of the inverse of time-evolution operators in symmetry relations prevents us from straightforwardly identifying the symmetries. To overcome this difficulty, a concept of $symmetry\ time\ frame$ has been introduced\cite{qw14}. The symmetry time frame requires a redefinition of the time-evolution operator by shifting the origin of time so that the time-evolution operator exhibits symmetric order of elemental operators in the time direction. In the case of $\tilde{U}(k)$ in Eq. (\ref{utk}), the redefined time-evolution operator $\tilde{U}'(k)$ fitted in the symmetric time frame is written down as 
\begin{align}
\tilde{U}'(k)=\tilde{C}(\theta_{1}/2)\,\tilde{S}(k)\,\tilde{\Phi}\,\tilde{G}\,\tilde{C}(\theta_{2})\,\tilde{G}^{-1}\,\tilde{\Phi}\,\tilde{S}(k)\,\tilde{C}(\theta_{1}/2),
\label{utpk}
\end{align}
which we can obtain by the unitary transformation; $\tilde{U}'(k)=e^{i\frac{\theta_{1}}{2}\sigma_{1}}\tilde{U}(k)e^{-i\frac{\theta_{1}}{2}\sigma_{1}}$. Here, we use the commutative property between operators $\tilde{G}$, $\tilde{S}(k)$, and $\tilde{\Phi}$ as they are described by exponentials of $\sigma_{3}$. By substituting $\tilde{U}'(k)$ in Eq. (\ref{utpk}) into Eqs. (\ref{pu})-(\ref{ptu}), we obtain conditions for elemental operators $\tilde{C}(\theta_{i})$, $\tilde{G}$, $\tilde{S}(k)$, and $\tilde{\Phi}$ to retain each symmetry. For example, in the case of TRS, we obtain the following two equations from left and right hand sides of Eq. (\ref{tu}) by substituting Eq. (\ref{utpk}):
\begin{align}
&\text{LHS}=[\tilde{\mT}\tilde{C}(\theta_{1}/2)\tilde{\mT}^{-1}][\tilde{\mT}\tilde{S}(k)\tilde{\mT}^{-1}][\tilde{\mT}\tilde{\Phi}\tilde{\mT}^{-1}][\tilde{\mT}\tilde{G}\tilde{\mT}^{-1}]\cdots,\nonumber\\
&\text{RHS}=[\tilde{C}^{-1}(\theta_{1}/2)][\tilde{S}^{-1}(-k)][\tilde{\Phi}^{-1}][\tilde{G}]\cdots.\nonumber
\end{align}
Comparing two equations, we obtain conditions for the elemental
operators, such as, $\tilde{T}\tilde{C}(\theta_{1})\tilde{T}^{-1}
= \tilde{C}^{-1}(\theta_{1})$, and so on. We summarize conditions on all elemental operators for various symmetries in Table \ref{tb:s}. Using Table \ref{tb:s}, we discuss symmetries of the time-evolution operator by starting from the unitary case, then including the gain-loss and phase operators step by step.

In the case $\gamma=\phi=0$: In this case, the time-evolution operator $\tilde{U}'(k)$ describes the unitary quantum walk and we consider conditions only on $\tilde{C}(\theta_{i})$ and $\tilde{S}(k)$ in Table \ref{tb:s}. From the anti-commutation relations of Pauli matrices, we identify that $\tilde{U}'(k)$ satisfies parity symmetry and TRS with the following symmetry operators:
\begin{align}
\tilde{\mP}=\sigma_{1},\ \tilde{\mT}=\sigma_{1}\mathcal{K}.
\label{pandtk}
\end{align}
Therefore, by combing the two symmetry operators in Eq. (\ref{pandtk}), the \PT symmetry operator is determined as
\begin{align}
\tilde{\mP}\tilde{\mT}=\sigma_{0}\mathcal{K},
\label{ptk}
\end{align}
where $\sigma_{0}=\text{diag}(1,1)$, and $\tilde{U}'(k)$ also possesses \PT symmetry. 

In the case $\gamma\neq0$ and $\phi=0$: The finite $\gamma$ makes
$\tilde{U}'(k)$ the non-unitary time-evolution operator and we should
consider the additional condition on the gain-loss operator $\tilde{G}$
as well as those on $\tilde{C}(\theta_{i})$ and $\tilde{S}(k)$ in Table
\ref{tb:s}. Since conditions on $\tilde{G}$ for parity
symmetry and TRS by symmetry operators in Eq. (\ref{pandtk}) are not
satisfied, the time-evolution operator $\tilde{U}'(k)$ has neither parity symmetry nor TRS. However, when we consider \PT symmetry, the condition $(\tilde{\mP}\tilde{\mT})\tilde{G}(\tilde{\mP}\tilde{\mT})^{-1}=\tilde{G}$ with $\tilde{\mP}\tilde{\mT}$ in Eq. (\ref{ptk}) is satisfied. Therefore, we identify \PT symmetry and confirm that the non-unitary time-evolution operator $\tilde{U}'(k)$ (with $\phi=0$) preserves \PT symmetry.

In the case $\gamma\neq0$ and $\phi\neq0$: Now, the condition on the phase operator in Table \ref{tb:s} is also maintained to retain \PT symmetry. We easily confirm the condition $(\tilde{\mP}\tilde{\mT})\tilde{\Phi}(\tilde{\mP}\tilde{\mT})^{-1}=\tilde{\Phi}^{*}$ with $\tilde{\mP}\tilde{\mT}$ in Eq. (\ref{ptk}). Thereby, we conclude that, nevertheless individual parity symmetry and TRS are broken in the non-unitary quantum walk with the phase operator in the homogeneous system, there presents \PT symmetry. 

We recall that the sufficient condition for quasi-energy being real
requires the other condition, namely,
the eigenvector of the non-unitary time-evolution operator is also one of the \PT symmetry operator. To check this, applying the unitary transformation $e^{i (\theta_1/2) \sigma_1 }$ to the eigenvector of $\tilde{U}(k)$ in Eq.\ (\ref{eq:eigenvector}), the eigenvector of $\tilde{U}^\prime(k)$ fitted in the symmetry time frame is described as $\ket{\Psi^\prime_{k,\pm}} = e^{i (\theta_1/2)\sigma_1}\ket{\Psi_{k,\pm}}$. Then, we can straightforwardly confirm the equation,
\begin{equation*}
 \tilde{\mP} \tilde{\mT} \ket{\Psi^\prime_{k,\pm}} = \pm e^{+i2\eta_k} \ket{\Psi^\prime_{k,\pm}},
\end{equation*}
as long as $\xi_k$ is real (then $\varepsilon$ is also real). Therefore, we confirm that the entirely real quasi-energy in Eq. (\ref{qe}) originates to \PT symmetry of the non-unitary time-evolution operator.

\begin{table*}[thbp]
\begin{center}
\caption{A list of conditions for elemental operators so that the
 time-evolution operator $\tilde{U}'(k)$ satisfies parity,
 time-reversal, and \PT, chiral, particle-hole, and parity-chiral
 symmetries. The first column indicates each symmetry and the second
 column represents the symmetry operators $\tilde{X}=\tilde{\mP},\,\tilde{\mT},\,\tilde{\mP}\tilde{\mT},\,\tilde{\Gamma},\,\tilde{\Xi}$ and $\tilde{\mP}\tilde{\Gamma}$, and the third column $\tilde{X}_{u}$ shows specific forms of symmetry operators which are derived from the unitary time-evolution operator with $\gamma=\phi=0$. The forth to seventh columns show conditions for the elemental operators to satisfy each symmetry. This table in this part should be read, $i.e.$, in order to satisfy parity symmetry the coin operator should satisfy $\tilde{\mP}\tilde{C}(\theta_{i})\tilde{\mP}^{-1}=\tilde{C}(\theta_{i})$. The  $\text{yes}$ or $\text{no}$, next to each condition explains the condition is satisfied or not, respectively, with the symmetry operator $\tilde{X}_{u}$. Note that $\tilde{C}(\theta_{i})=e^{i\theta_{i}\sigma_{1}},\,\tilde{S}(k)=e^{ik\sigma_{3}},\,\tilde{G}=e^{\gamma\sigma_{3}},$ and $\tilde{\Phi}=e^{i\phi\sigma_{3}}$. We use the following relations; $\tilde{C}^{-1}(\theta_{i})=\tilde{C}(-\theta_{i}),\,\tilde{S}^{-1}(k)=\tilde{S}(-k),\,\tilde{S}^{-1}(-k)=\tilde{S}(+k),$ and $\tilde{\Phi}^{-1}=\tilde{\Phi}^{\ast}$.}
\vspace{5mm}
\begin{tabular}{lllllll}
\hline\hline
 Symmetry & $\tilde{X}$\hspace{3mm} & $\tilde{X}_{u}$\hspace{5mm} & $\tilde{X}\tilde{C}(\theta_{i})\tilde{X}^{-1}$\hspace{4mm} & $\tilde{X}\tilde{S}(k)\tilde{X}^{-1}$\hspace{4mm} & $\tilde{X}\tilde{G}\tilde{X}^{-1}$\hspace{4mm} & $\tilde{X}\tilde{\Phi}\tilde{X}^{-1}$\hspace{4mm} \\ \hline
Parity symmetry & $\tilde{\mP}$ & $\sigma_{1}$ & $\tilde{C}(+\theta_{i})\ [\text{yes}]$ & $\tilde{S}(-k)\ [\text{yes}]$ & $\tilde{G}\ [\text{no}]$ & $\tilde{\Phi}\hspace{2.6mm} [\text{no}]$  \\
Time-reversal symmetry (TRS) & $\tilde{\mT}$ & $\sigma_{1}\mK$ & $\tilde{C}(-\theta_{i})\ [\text{yes}]$ & $\tilde{S}(+k)\ [\text{yes}]$ & $\tilde{G}\ [\text{no}]$ & $\tilde{\Phi}^{*}\ [\text{no}]$  \\
\PT symmetry & $\tilde{\mP}\tilde{\mT}$ & $\sigma_{0}\mK$ & $\tilde{C}(-\theta_{i})\ [\text{yes}]$ & $\tilde{S}(-k)\ [\text{yes}]$ & $\tilde{G}\ [\text{yes}]$ & $\tilde{\Phi}^{*}\ [\text{yes}]$  \\ \hline
Chiral symmetry & $\tilde{\Gamma}$ & $i\sigma_{2}$ & $\tilde{C}(-\theta_{i})\ [\text{yes}]$ & $\tilde{S}(-k)\ [\text{yes}]$ & $\tilde{G}\ [\text{no}]$ & $\tilde{\Phi}^{*}\ [\text{yes}]$  \\
Particle-hole symmetry (PHS) & $\tilde{\Xi}$ & $\sigma_{3}\mK$ & $\tilde{C}(+\theta_{i})\ [\text{yes}]$ & $\tilde{S}(-k)\ [\text{yes}]$ & $\tilde{G}\ [\text{yes}]$ & $\tilde{\Phi}\hspace{2.6mm} [\text{no}]$  \\
Parity-chiral symmetry (PCS) & $\tilde{\mP}\tilde{\Gamma}$ & $\sigma_{3}$ & $\tilde{C}(-\theta_{i})\ [\text{yes}]$ & $\tilde{S}(+k)\ [\text{yes}]$ & $\tilde{G}\ [\text{yes}]$ & $\tilde{\Phi}^{*}\ [\text{no}]$  \\ \hline\hline
\label{tb:s}
\end{tabular} 
\end{center}
\end{table*}

\subsubsection{extra symmetries} 
The time-evolution operator of the non-unitary quantum walk in Eq. (\ref{utpk}) can posses extra symmetries. Here, we discuss such symmetries which are intensively studied for topological phases of the quantum walk\cite{qw10,qw11,qw12,qw13,qw14,qw15}. These extra symmetries are chiral symmetry and particle-hole symmetry (PHS) defined for a Hamiltonian $H$ as
\begin{align}
&\Gamma\, H\, \Gamma^{-1}=-H,
\label{ch}\\
&\Xi\, H\, \Xi^{-1}=-H,
\label{phh}
\end{align}
respectively. The chiral symmetry operator $\Gamma$ is a unitary operator, while the PHS operator $\Xi$ is an anti-unitary one. These two symmetries guarantee that the system has a pair of eigenstates with opposite sign of eigenvalues if the eigenvalue is real. Accordingly, eigenenergy appears symmetric with respect to zero energy. Following the same procedure with before, we convert Eqs. (\ref{ch}) and (\ref{phh}) to symmetry relations for the time-evolution operator:
\begin{align*}
\Gamma\, U\, \Gamma^{-1} &= U^{-1},\\
\Xi\, U\, \Xi^{-1} &= U.
\end{align*}
Defining the symmetry operators as
\begin{align*}
\Gamma &=\sum_n \ket{n}\bra{n}\otimes\tilde{\Gamma}=\sum_k \ket{k}\bra{k}\otimes\tilde{\Gamma},\\
\Xi &=\sum_n \ket{n}\bra{n}\otimes\tilde{\Xi}=\sum_k \ket{-k}\bra{k}\otimes\tilde{\Xi},
\end{align*}
we derive relations to retain chiral symmetry and PHS:
\begin{align}
&\tilde{\Gamma}\, \tilde{U}(k)\, \tilde{\Gamma}^{-1}=\tilde{U}^{-1}(+k),
\label{cu}\\
&\tilde{\Xi}\, \tilde{U}(k)\, \tilde{\Xi}^{-1}=\tilde{U}(-k).
\label{phu}
\end{align}
Substituting Eq. (\ref{utpk}) into Eqs. (\ref{cu}) and (\ref{phu}), we again obtain conditions on the elemental operators to retain chiral symmetry and PHS as shown in Table \ref{tb:s}. Due to $2\pi$ periodicity of the quasi-energy, if the time-evolution operator satisfies Eq. (\ref{cu}) and/or (\ref{phu}), the quasi-energy appears symmetric with respect to $\varepsilon=0$ and $\pi$.

In the case $\gamma=\phi=0$: At first, we focus on conditions on the
coin and shift operators in the case of chiral symmetry in Table
\ref{tb:s} for this unitary quantum walk. We find that, with the symmetry operator $\tilde{\Gamma}=i\sigma_{2}$, chiral symmetry is retained. It is known that if TRS and chiral symmetry are presented, PHS is simultaneously retained with the symmetry operator $\tilde{\Xi}=\tilde{\Gamma}\tilde{\mT}$. In summary, by using 
\begin{align}
\tilde{\Gamma}=i\sigma_{2},\ \tilde{\Xi}=\sigma_{3}\mathcal{K},
\label{candphk}
\end{align}
the unitary time-evolution operator $\tilde{U}'(k)$ has extra symmetries, chiral symmetry and PHS.

In the case $\gamma\neq0$ and $\phi=0$: In order to retain chiral
symmetry and PHS for this non-unitary quantum walk, the gain-loss
operator $\tilde{G}$ should be unchanged ($\tilde{X}\tilde{G}\tilde{X}^{-1} =
\tilde{G}$) when $\tilde{X}=\tilde{\Gamma}$ or
$\tilde{\Xi}$ in Eq. (\ref{candphk}) is acted on. We understand that
$\tilde{X}=\tilde{\Xi}$ keeps $\tilde{G}$ as is, while $\tilde{X}=\tilde{\Gamma}$ does not. Thereby, only PHS survives after including gain and loss effects. However, we can introduce a new symmetry combined with parity and chiral symmetries,
\begin{equation*}
(\mP \Gamma)\, U\, (\mP \Gamma)^{-1}= U^{-1},
\end{equation*}
that we call parity-chiral symmetry (PCS). Taking account of Eqs. (\ref{pu}) and (\ref{cu}), we derive the symmetry relation for PCS
\begin{align}
(\tilde{\mP}\tilde{\Gamma})\, \tilde{U}(k)\, (\tilde{\mP}\tilde{\Gamma})^{-1}=\tilde{U}^{-1}(-k),
\label{pcu}
\end{align}
and then obtain conditions on each elemental operator as listed in Table
\ref{tb:s}. We note that PCS also guarantees the symmetric behavior of
the quasi-energy with respect to $\varepsilon=0$ and $\pi$. From
Eqs. (\ref{pandtk}) and (\ref{candphk}), the PCS operator becomes 
\begin{align}
\tilde{\mP}\tilde{\Gamma}=\sigma_{3},
\label{pck}
\end{align}
(we ignore an unimportant minus sign). With the above symmetry operator $\tilde{\mP}\tilde{\Gamma}$, we confirm that $\tilde{U}'(k)$ possesses PCS, and the symmetric property of the quasi-energy is guaranteed by PHS and PCS.

In the case $\gamma\neq0$ and $\phi\neq0$: Finally, we consider the
non-unitary quantum walk with finite phases whose quasi-energy is given in
Eq. (\ref{qe}). To retain PHS and PCS, the phase operator should satisfy
$\tilde{\Xi}\tilde{\Phi}\tilde{\Xi}^{-1}=\tilde{\Phi}$ and
$(\tilde{\mP}\tilde{\Gamma})\tilde{\Phi}(\tilde{\mP}\tilde{\Gamma})^{-1}=\tilde{\Phi}^{*}$,
respectively. However, both conditions are not satisfied with the symmetry operators in Eqs. (\ref{candphk}) and (\ref{pck}). Thereby, the finite $\gamma$ and $\phi$ break all symmetries which guarantee a pair of eigenstates with the opposite quasi-energy. 

While the above result implies that a pair of quasi-energy in
Eq. (\ref{qe}) does not originate to symmetry, we can still find out
contributions of symmetry by introducing a modified version of
parity symmetry defined below. Because of translation symmetry in the
homogeneous system, we re-express the
time-evolution operator in Eq. (\ref{utpk}) by including the phase operator into the shift operator as
\begin{align}
\tilde{U}'(k)=\tilde{C}(\theta_{1}/2)\,\tilde{S}(k+\phi)\,\tilde{G}\,\tilde{C}(\theta_{2})\,\tilde{G}^{-1}\,\tilde{S}(k+\phi)\,\tilde{C}(\theta_{1}/2).
\label{mutpk}
\end{align}
Next, we introduce the modified parity symmetry operator with phase modulations defined as
\begin{align}
\mP_{\phi}=\sum_{n}e^{-i2\phi n}\ket{-n}\bra{n}\otimes\tilde{\mP}_{\phi}
=\sum_{k}\ket{-k-2\phi}\bra{k}\otimes\tilde{\mP}_{\phi}.\nonumber
\end{align}
By combing the modified parity symmetry operator $\mP_{\phi}$ and chiral symmetry operator $\Gamma$, the condition on the shift operator $\tilde{S}(k+\phi)$ to retain modified PCS, $(\tilde{\mP}_{\phi}\tilde{\Gamma})\tilde{U}^\prime(k)(\tilde{\mP}_{\phi}\tilde{\Gamma})^{-1}=\tilde{U}^{\prime-1}(-k-2\phi)$, becomes 
\begin{align}
(\tilde{\mP}_{\phi}\tilde{\Gamma})\tilde{S}(k+\phi)(\tilde{\mP}_{\phi}\tilde{\Gamma})^{-1}=\tilde{S}(k+\phi).\nonumber
\end{align}
which is satisfied by the symmetry operator $\tilde{\mP}_\phi
\tilde{\Gamma}=\sigma_3$. Note that conditions for $\tilde{C}(\theta_{i})$ and $\tilde{G}$ to retain modified PCS are the same with those of PCS, since both operators are $k$ independent. Thereby, we identify that a pair of quasi-energy in Eq. (\ref{qe}) originates from modified PCS.

\subsection{Symmetries in inhomogeneous systems}
\label{sec:siis}
\begin{figure}[t]
\begin{center}
\includegraphics[scale=0.25]{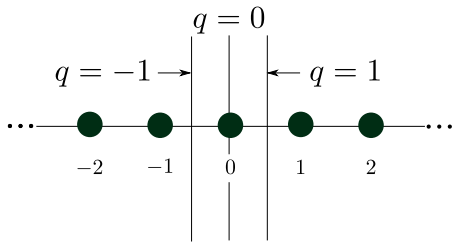}
\caption{\ColorOnline The difference of the reflection points of the parity symmetry operator. When $q=0$, the reflection point is on the site $n=0$. When $q=\pm1$, the reflection point is between sites $n=0$ and $n=\pm1$.}
\label{fig:parity}
\end{center}
\end{figure}

Next, we consider \PT symmetry, PHS, and PCS of the time-evolution
operator of the non-unitary quantum walk in Eq.\ (\ref{uhx}) with position dependent
parameters. Therefore, we need to consider the time-evolution operator
in the position representation. Taking the symmetry operators for
internal space in Eqs. (\ref{ptk}), (\ref{candphk}), and (\ref{pck}) into account, those in the position representation are described as
\begin{subequations}
\begin{align}
\mPT&=\sum_{n}\ket{-n+q}\bra{n}\otimes\sigma_{0}\mK,
\label{ptx}\\
\Xi&=\sum_{n}\ket{n}\bra{n}\otimes\sigma_{3}\mK,
\label{phx}\\
\mP\Gamma&=\sum_{n}\ket{-n+q}\bra{n}\otimes\sigma_{3},
\label{pcx}
\end{align}
\label{eq:symmetry op inhomogeneous}\end{subequations}
where the index $q$ is introduced to determine the origin of the space
reflection point (see Fig. \ref{fig:parity}) because we treat lattice
systems.  
By using the symmetry operators in Eqs. (\ref{ptx})-(\ref{pcx}), each symmetry defined for the time-evolution operator in the position representation becomes
\begin{subequations}
\begin{align}
(\mPT)U(\mPT)^{-1}&=U^{-1},
\label{pthx}\\
\Xi U\Xi^{-1}&=U,
\label{phhx}\\
(\mP\Gamma)U(\mP\Gamma)^{-1}&=U^{-1}.
\label{pchx}
\end{align}
\label{eq:symmetry inhomogeneous}\end{subequations}
Equations (\ref{eq:symmetry op inhomogeneous}) and (\ref{eq:symmetry inhomogeneous}) guarantee that if two of the above three symmetries are confirmed, there also exists the other symmetry which is derived by combining the confirmed two symmetries. Even in the position representation, we need to use the time-evolution operator fitted into the symmetry time frame written as
\begin{align}
U^\prime=C(\theta_{1}/2)\,S\,G_{2}\,\Phi_{2}\,C(\theta_{2})\,S\,G_{1}\,\Phi_{1}\,C(\theta_{1}/2).
\label{uhpx}
\end{align}

As shown in Sec. \ref{sec:sihs}, when parameters of the coin, gain-loss,
and phase operators are position independent, conditions to retain each
symmetry are reduced to conditions to the elemental operators as
summarized in Table \ref{tb:s}. This simplification is based on the fact
that all of the operators $\tilde{G}$, $\tilde{S}(k)$, and
$\tilde{\Phi}$ are described by exponentials of $\sigma_{3}$, and then
they are commutative. However, when the parameters depend on position,
the shift operator $S$ is not commutative with gain-loss operator
$G_{i}$ and phase operator $\Phi_{i}$. Thus, we need to consider
conditions for operators $SG_{i}\Phi_{i}$ as a whole. For example, the condition to retain \PT symmetry for the time-evolution operator is derived as follows. By substituting Eq. (\ref{uhpx}) into Eq. (\ref{pthx}), the left and right hand sides become
\begin{align}
&\text{LHS}=[(\mPT)\,C(\theta_{1}/2)\,(\mPT)^{-1}]\,
[(\mPT)\,SG_{2}\Phi_{2}\,(\mPT)^{-1}]\cdots,\nonumber\\
&\text{RHS}=[C^{-1}(\theta_{1}/2)]\,[(S G_{1}\Phi_{1})^{-1}]\cdots,\nonumber
\end{align}
respectively. By comparing these two equations, we obtain the conditions to retain \PT symmetry for the time-evolution operator of the non-unitary quantum walk in inhomogeneous systems as
\begin{subequations}
\begin{align}
(\mPT)C(\theta_{i})(\mPT)^{-1}&=C^{-1}(\theta_{i}),
\label{ptcx}\\
(\mPT)(SG_{i}\Phi_{i})(\mPT)^{-1}
&=(SG_{j}\Phi_{j})^{-1},
\label{ptgspx}
\end{align}
\label{ptcgspx}\end{subequations}
where $i,j=1,2$ and $i\neq j$. From Eq. (\ref{ptcgspx}), we obtain conditions imposed on each position dependent parameter to retain \PT symmetry as 
\begin{subequations}
\begin{align}
\theta_{i}(n)&=\theta_{i}(-n+q),
\label{pttx}\\
g_{1,L}(n)&=[g_{2,L}(-n+q+1)]^{-1},
\label{ptglx}\\
g_{1,R}(n)&=[g_{2,R}(-n+q-1)]^{-1},
\label{ptgrx}\\
\phi_{1,L}(n)&=\phi_{2,L}(-n+q+1),
\label{ptplx}\\
\phi_{1,R}(n)&=\phi_{2,R}(-n+q-1).
\label{ptprx}
\end{align}
\label{ptpx}\end{subequations}
We find that, on one hand, the parameter $\theta_{i}(n)$ of the coin
operator is uncorrelated in time direction, which means that, $\theta_1
(n)$ and $\theta_2 (n)$ can be determined independently. On the other
hand, parameters of gain-loss and phase operators have strict
restrictions in time direction as well as in position
space. We note that when
conditions in Eqs. (\ref{ptglx}) and (\ref{ptgrx}) are satisfied, the absolute value of the determinant of the time-evolution operator $U$ in inhomogeneous systems remains to be one, even though the determinant of each $G_{i}$ is not one. We should also remind that, while the conditions Eq.\ (\ref{ptpx}) guarantee that the time-evolution operator has \PT symmetry, they do not guarantee that eigenvectors of the time-evolution operator are those of the \PT symmetry operator.

In the same way, we can obtain conditions to preserve PCS and
PHS for the time-evolution operator in inhomogeneous
systems.  We find that PCS is maintained under the following conditions:
\begin{subequations}
\begin{align}
\theta_{i}(n)&=\theta_{i}(-n+q),
\label{pctx}\\
g_{1,L}(n)&=[g_{2,L}(-n+q+1)]^{-1},
\label{pcglx}\\
g_{1,R}(n)&=[g_{2,R}(-n+q-1)]^{-1},
\label{pcgrx}\\
\phi_{1,L}(n)&=-\phi_{2,L}(-n+q+1),
\label{pcprx}\\
\phi_{1,R}(n)&=-\phi_{2,R}(-n+q-1).
\label{pcplx}
\end{align}
\label{pcpx}\end{subequations}
Comparing the above conditions, Eq.\ (\ref{pcpx}),  with those for \PT symmetry in Eq.\ (\ref{ptpx}), we understand that, while Eqs.\ (\ref{pctx})-(\ref{pcgrx}) are the same with Eqs.\ (\ref{pttx})-(\ref{ptgrx}), the conditions on phases $\phi_{i,\sigma}(n)$ to retain \PT symmetry and PCS cannot be simultaneously satisfied unless $\phi_{i,\sigma}(n)=0$.
This gives another conclusion that PHS is retained only if $\phi_{i,\sigma}(n)=0$ since PHS can be defined as the combination of \PT symmetry and PCS, $\Xi = (\mPT)\, (\mP \Gamma)$. By combining Eqs.\ (\ref{ptpx}) and
(\ref{pcpx}), we also understand that there is no constraint on
$\theta_i(n)$ and $g_{i,\sigma}(n)$ to retain PHS.

\section{applications}
\label{sec:applications}
Finally, we apply results to retain various symmetries obtained in Sec.\ \ref{sec:paes} into specific models of non-unitary quantum walks. At first, we identify symmetries of the non-unitary quantum walk realized in the experiment\cite{regb1}. Secondly, we show the numerical results of walker's time-evolution in the homogeneous system consider in Sec. \ref{sec:sihs}. For the other example, we demonstrate that, for an inhomogeneous non-unitary quantum walk where four distinct spatial regions exist, the time-evolution operator possesses \PT symmetry and the quasi-energy becomes entirely real.
\subsection{Symmetries satisfied in the experiment}
\label{sec:experiment}
Here, we directly identify symmetries of the non-unitary quantum walk realized in the experiment\cite{regb1} from the time-evolution operator. The time-evolution operator in the experiment, $U_{\text{ex}}$, is given by Eq. (\ref{uhx}) by assigning the following parameters:
\begin{subequations}
\begin{align}
\theta_{1}(n)&=\theta_{2}(n)=\pi/4,
\label{t12e}\\
g_{1,L}(n)&=[g_{2,L}(n)]^{-1}=e^{+\gamma_0},
\label{gle}\\
g_{1,R}(n)&=[g_{2,R}(n)]^{-1}=e^{-\gamma_0},
\label{gre}\\
\phi_{1,L}(n)&=\phi_{2,L}(n)=0,
\label{ple}\\
\phi_{1,R}(n)&=\phi_{2,R}(n)=\left\{\begin{array}{ll}
-\phi_{0}&\text{for mod}(n+3,4)=1,\,2,\\
+\phi_{0}&\text{for mod}(n+3,4)=3,\,0.
\end{array}\right.
\label{pre}
\end{align}
\label{pex}\end{subequations}
The quasi-energy of this time-evolution operator becomes
\begin{align}
\cos(\pm\varepsilon)=-\frac{1}{2}\cos\phi_0\cosh(2\gamma_0)\pm\sqrt{f_k(\gamma_0,\phi_0)},
\label{qee}
\end{align}
where
\begin{align*}
f_k(\gamma_0,\phi_0)=\frac{1}{8}[&\cosh(4\gamma_0)(\cos^{2}\phi_0-1)\\
&-3\cos^{2}\phi_0+4+\cos k].
\end{align*}
Regarding \PT symmetry, we can confirm that all parameters in Eq. (\ref{pex}) satisfy conditions in Eq. (\ref{ptpx}) to retain \PT symmetry, especially, by choosing $q=-1$ for $\phi_{i,L}(n)$ which only depends on the position. Therefore, we can identify \PT symmetry of the non-unitary time-evolution operator $U_{\text{ex}}$ with the symmetry operator in Eq. (\ref{ptx}).

\begin{figure}[t]
\begin{center}
\includegraphics[width=8.0cm]{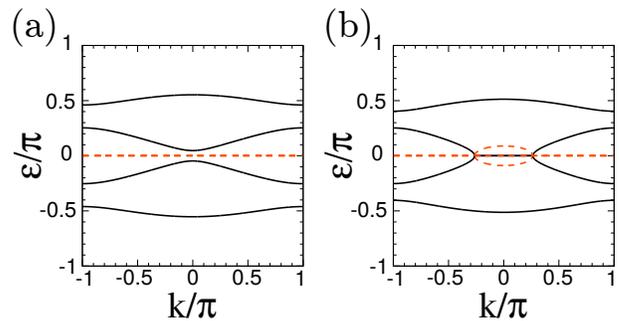}
\caption{\ColorOnline The quasi-energy as a function of $k$ in
 Eq. (\ref{qee}) with various gain-loss parameters when $\phi_{0}=6\pi/5$. The solid (dashed) curve represents the real (imaginary) part of the quasi-energy. (a)When $e^{\gamma}=1.1$, the quasi-energy is entirely real. (b)When $e^{\gamma}=1.4$, a part of the quasi-energy becomes complex. In both cases, quasi-energy exists being symmetric with respect to $\varepsilon=0$.}
\label{fig:dex}
\end{center}
\end{figure}

From Eq. (\ref{qee}) and Fig. \ref{fig:dex}, we expect that the time-evolution operator $U_{\text{ex}}$ also has PHS and PCS because there appear pairs with the opposite quasi-energy $\pm\varepsilon$. However, as shown in Sec. \ref{sec:siis}, the finite $\phi_{i,\sigma}(n)$ prevents PHS and PCS. This problem is solved by introducing a modified PHS operator with a position shift by $r$ as
\begin{align}
\Xi_{r}=\sum_{n}\ket{n+r}\bra{n}\otimes\sigma_{3}\mathcal{K}
\label{mphpx}
\end{align}
By using the modified PHS operator $\Xi_{r}$, the condition on the phase parameter to satisfy $\Xi_{r}U\Xi_{r}^{-1}=U$ is derived as 
\begin{align}
\phi_{i,\sigma}(n)=-\phi_{i,\sigma}(n+r).
\label{mphx}
\end{align}
Inputting $r=2$, we confirm that the phase parameter in Eq. (\ref{pre}) satisfies Eq. (\ref{mphx}). Therefore, the time-evolution operator $U_{\text{ex}}$ also preserves modified PHS.

\subsection{Time-evolution of probability distributions of homogeneous non-unitary quantum walks}
\label{sec:time-evolution}
\begin{figure*}[thbp]
\begin{center}
\includegraphics[width=17.5cm]{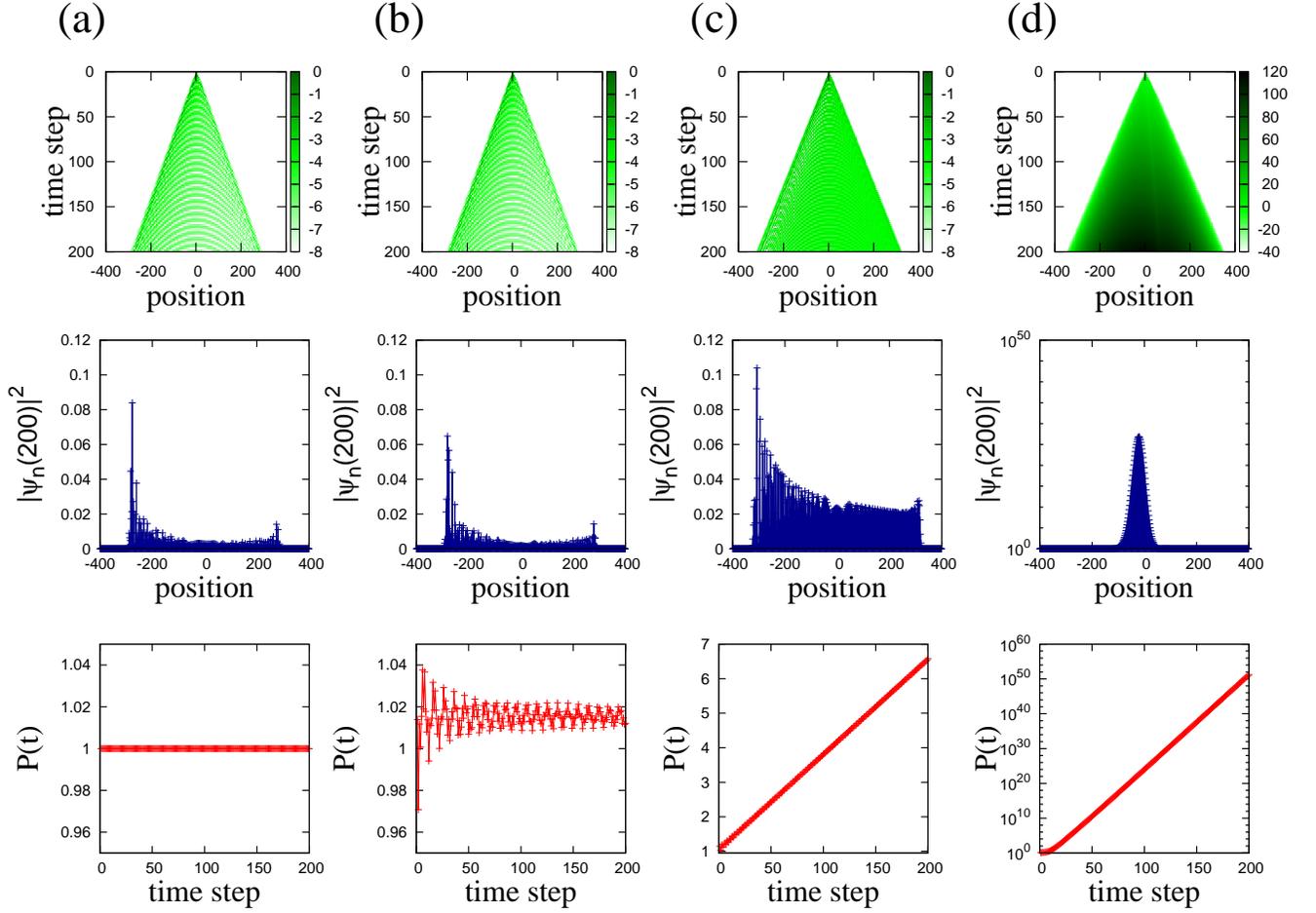}
\caption{\ColorOnline The time-evolutions for the quantum walk in
 the homogeneous system with various gain-loss parameters; (a) $e^{\gamma}=1$ (the unitary quantum walk), (b) $e^{\gamma}=1.1$ (the non-unitary quantum walk with entirely real quasi-energy) , (c) $e^{\gamma}=1.34\cdots$ (the non-unitary quantum walk at the exceptional point), and (d) $e^{\gamma}=1.5$ (the non-unitary quantum walk with complex quasi-energy). The other parameters  $\theta_1=\pi/4$, $\theta_2=-\pi/7$, $\phi=0$, and the initial state $\ket{\psi(0)}=\ket{0}\otimes\ket{R}$ are used for all cases (a)-(d).
(top panels) The contour maps of the logarithm of the probability distribution 
$\ln(|\psi_{n}(t)|^{2})$ in the position- and time plain.
(middle panels) The probability distributions after 200 time steps $|\psi_{n}(t=200)|^{2}$. (bottom panels) The time step dependence of the sum of the probability distributions $P(t)$.}
\label{fig:tehomogeneous}
\end{center}
\end{figure*}
Next, we numerically demonstrate the time evolution of probability distributions of non-unitary quantum walks in homogeneous systems. To this end, we employ the time-evolution operator in Eq.\ (\ref{uk}). We note that we define the probability distribution at a position $n$ at a time $t$ as
\begin{equation*}
|\psi_{n}(t)|^{2}=|\psi_{n,L}(t)|^{2}+|\psi_{n,R}(t)|^{2} 
\end{equation*}
even for non-unitary quantum walks although, in non-Hermitian quantum mechanics, the biorthogonality of eigenvectors (of a Hamiltonian or time-evolution operator) should be taken into account for normalized inner products. Because of this, the sum of the probability distributions over the position space
\begin{equation*}
 P(t)=\sum_{n} |\psi_{n}(t)|^{2}
\end{equation*}
need not to be one for the non-unitary quantum walk, while $P(t)=1$ for the unitary quantum walk. This choice stems from the fact that the quantity
$|\psi_{n}(t)|^{2}$ calculated numerically agrees well with the intensity distribution of laser pulses observed experimentally in the optical-fibre loops with loss as reported in Ref.\ \cite{regb4}. 

In Fig.\ \ref{fig:tehomogeneous}, we show numerical results on the time-evolution for the homogeneous quantum walk in Eq.\ (\ref{uk}).
The parameters are the same with the parameter set in Fig.\ \ref{fig:qe}, namely, (a) $e^{\gamma}=1$ (the unitary quantum walk), (b) $e^{\gamma}=1.1$ (the non-unitary quantum walk with entirely real quasi-energy), (c)
$e^{\gamma}=1.34\cdots$ (the non-unitary quantum walk at the exceptional point), and (d) $e^{\gamma}=1.5$ (the non-unitary quantum walk with complex
quasi-energy). Comparing the probability distributions in Figs.\
\ref{fig:tehomogeneous} (a) and (b), when the non-unitary quantum walk has entirely real quasi-energy, the time-evolution is not largely different from that of the unitary quantum walk. One exception is that the sum of the probability distribution $P(t)$ exhibits tiny oscillations around $P(t)\approx 1$ with time in the non-unitary case [Fig.\ \ref{fig:tehomogeneous} (b-bottom)], while $P(t)=1$ in the unitary quantum walk [Fig.\ \ref{fig:tehomogeneous} (a-bottom)]. 

However, as increasing $\gamma$ further, the time evolution of the
non-unitary quantum walk drastically changes. At the exceptional point, the sum of the probability distribution $P(t)$ grows linearly with time as shown in Fig.\ \ref{fig:tehomogeneous} (c-bottom), and when part of quasi-energies become complex, $P(t)$ grows exponentially with time as shown in Fig.\ \ref{fig:tehomogeneous} (d-bottom). Remarkably, in the latter case,  the probability distribution after 200 time step is well approximated by the Gaussian distribution [Fig.\ \ref{fig:tehomogeneous} (d-middle)], in contrast
with other cases (a)-(c). We note that linear and exponential grows of the sum of the probability distributions $P(t)$ are observed in Ref.\ \cite{regb1} under the different setup, and the Gaussian distribution of the probability distribution is also reported in Refs.\ \cite{Longhi10,regb4}. Therefore, these observations which are available by experiments can be considered as a manifestation of non-unitary time evolution.

\subsection{Non-unitary quantum walks with four distinct regions}
\label{sec:4phase}
\begin{figure*}[t]
\begin{center}
\includegraphics[width=16.5cm]{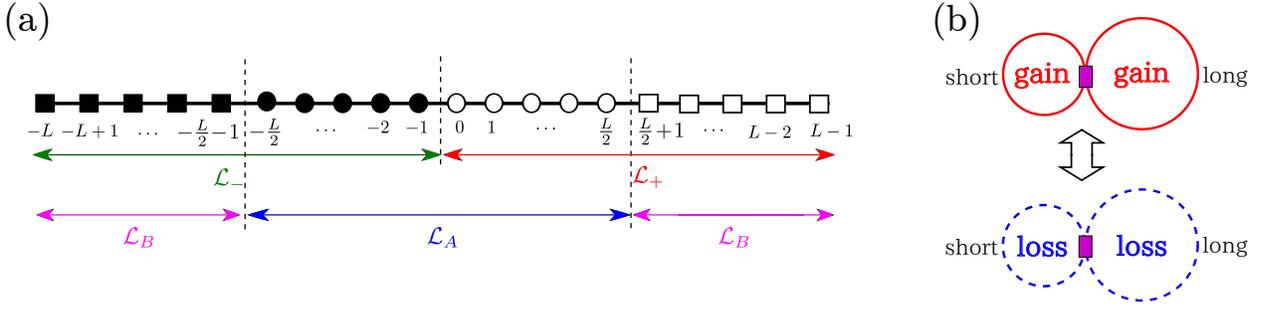}
\caption{\ColorOnline (a) A schematic view of the non-unitary quantum walk with four distinct spatial regions. (b)A schematic view to explain gain-loss operations in the experiment by the optical-fibre loops.}
\label{fig:4phase}
\end{center}
\end{figure*}

Although we can construct various time-evolution operators of non-unitary
quantum walks in inhomogeneous systems with \PT symmetry by employing
the conditions in Eq.\ (\ref{ptpx}), keeping real number of the quasi-energy requires the additional condition that eigenstates of the time-evolution operator are those of the \PT symmetry operator. Since it is our empirical fact that the additional condition is often broken in systems with strongly position dependent parameters, here we treat a rather moderate inhomogeneous non-unitary quantum walk as shown in Fig.\ \ref{fig:4phase} (a).
This system has four distinct spatial regions with different parameters by combinations of ${\cal
L}_{A/B}$ and ${\cal L}_{+/-}$ where each region are defined as
\begin{align*}
{\cal L}_A: &\quad -L/2 \le n \le L/2,\\
{\cal L}_B: &\quad n \le -L/2-1,\quad n \ge L/2+1,\\
{\cal L}_+: &\quad n \ge 0,\\
{\cal L}_-: &\quad n \le -1.
\end{align*}
Taking account of Eq.\ (\ref{ptpx}) with $q=0$, we choose parameters of the elemental operators as follows for instance:
\begin{subequations}
\begin{align}
\theta_{1}(n)&=\left\{
\begin{array}{ll}
+\pi/4\quad & n \in {\cal L}_{A},\\
-\pi/8 & n \in {\cal L}_{B},\\
\end{array}
\right.\\
\theta_{2}(n)&=\left\{
\begin{array}{ll}
-\pi/3\quad & n \in {\cal L}_{A},\\
+\pi/6 & n \in {\cal L}_{B},\\
\end{array}
\right.\\
g_{1,L}(n)&=[g_{2,L}(-n+1)]^{-1}=
\left\{
\begin{array}{ll}
1.1\quad & n \in {\cal L}_{-},\\
1.2 & n \in {\cal L}_{+},\\
\end{array}
\right.\\
g_{1,R}(n)&=[g_{2,R}(-n+1)]^{-1}=
\left\{
\begin{array}{ll}
1.2\quad & n \in {\cal L}_{-},\\
1.1 & n \in {\cal L}_{+},\\
\end{array}
\right.\\
\phi_{1,L}(n)&=\phi_{2,L}(-n+1)=
\left\{
\begin{array}{ll}
 \pi/4\quad & n \in {\cal L}_{-},\\
\pi/8 & n \in {\cal L}_{+},\\
\end{array}
\right.\\
\phi_{1,R}(n)&=\phi_{2,R}(-n+1)=
\left\{
\begin{array}{ll}
-\pi/3\quad & n \in {\cal L}_{-},\\
-\pi/6 & n \in {\cal L}_{+},\\
\end{array}
\right.
\end{align}
\label{eq:4phase}\end{subequations}
We emphasize that $\theta_i(n)$ is symmetric with respect to the origin of
position space, while $g_{i,\sigma}(n)$ and $\phi_{i,\sigma}(n)$ are
not. We also remark that the first (second) gain-loss operator $G_{1(2)}$ only amplifies (dumps) wave function amplitudes of both left and right mover
components as shown in Fig.\ {\ref{fig:4phase}} (b), in contrast to 
the experimental setup in Fig.\ {\ref{fig:es}} (a).

\begin{figure}[t]
\begin{center}
\includegraphics[width=6.0cm]{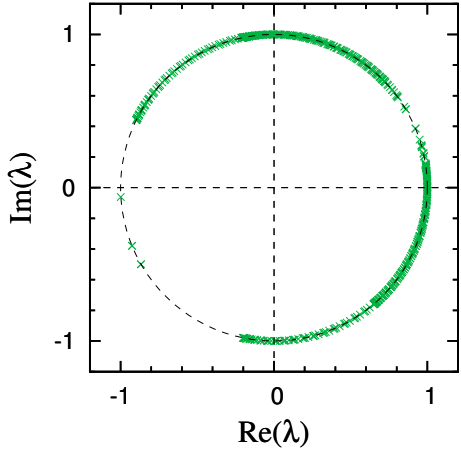}
\caption{\ColorOnline The eigenvalue $\lambda$ (green crossed) of the time-evolution operator of the non-unitary quantum walk with parameters in Eq.\ (\ref{eq:4phase}) plotted on a complex plain .}
\label{fig:4phase_spectrum}
\end{center}
\end{figure}

We numerically calculate eigenvalues of the time-evolution operator $U$ assigned the above parameters by imposing periodic boundary conditions to both ends $L-1$ and $-L$ with $L=128$. As shown in Fig.\ \ref{fig:4phase_spectrum} we clearly see that all eigenvalues stay on a unit circle in a complex plain, which indicates that the quasienergy is entirely real. Furthermore, eigenvalues are not symmetric with respect to
$\varepsilon=0,\pi$, because the position dependent phase parameters
$\phi_{i,\sigma}(n)$ break both PHS and PCS. 

\section{Summary and Discussion}
\label{sec:summary}
We have explicitly defined the \PT symmetry operator for the time evolution operator of the non-unitary quantum walk given in Eq.\ (\ref{uhx}), and identified necessary and sufficient conditions, Eq.\ (\ref{ptpx}), on position dependent parameters of the elemental operators to retain \PT symmetry. Taking
account of the conditions, we have succeeded to clarify the presence of
\PT symmetry of the non-unitary quantum walk realized in the experiment
by using optical-fibre loops\cite{regb1} from the time-evolution operator. This has been accomplished for the first time by employing the concept of the symmetry time frame which had been developed in the recent work on topological phases of quantum walks\cite{qw12}. At the same time, we have also studied extra symmetries embedded in the time evolution operator of the non-unitary quantum walk, such as chiral symmetry, PHS, PCS, and so on. In Sec.\ \ref{sec:time-evolution}, we have numerically demonstrated time-evolution of probability distributions for the homogeneous non-unitary quantum walk, and shown that those of the non-unitary quantum walk with entirely real quasi-energy are completely different from those with complex quasi-energy. Besides, we have also demonstrated in Sec. \ref{sec:4phase} that the inhomogeneous non-unitary quantum walk which has \PT symmetry and even possesses entirely real quasienergy is possible. 

We believe that the result obtained in the present work stimulates
further developments on \PT symmetry of non-unitary time-evolution
operators which has not yet been studied enough, compared with
non-Hermitian Hamiltonians. Also, the conditions Eq.\ (\ref{ptpx}) would
strongly support the experiment by using the optical-fibre loops\cite{regb1} as the versatile platform for studying phenomena originating to \PT symmetry. 
Besides this, although we have focused on the optical-fibre setup in the
present work, our result can be straightforwardly applied to other setups of the quantum walk. Furthermore, we can easily generalize our theory to the non-unitary quantum walk only with dissipation, which would be much easier to realized in various experimental setups. In addition, since we have shown that the non-unitary quantum walk can retain important symmetries to establish topological phases, it would be interesting to study topological phases and corresponding edge states of the non-unitary quantum walk, which we will report on elsewhere. 

An important open problem is to identify a generalized condition to retain real quasi-energy of the non-unitary quantum walk. According to progresses on \PT symmetry of non-Hermitian Hamiltonians, it is already known that the argument on \PT symmetry can be generalized as, if a Hamiltonian $H$ satisfies a pseudo-Hermiticity condition $\eta H \eta^{-1} = H^\dagger$ with a positive operator $\eta$ which may not be related to parity symmetry, eigenenergy could become real\cite{mostafazadeh02,mostafazadeh04}. Indeed, we observed possibly relating phenomena in our non-unitary quantum walk setup because quasi-energy becomes entirely real even when $\theta_1(n)$ is completely random in position space. This suggests a possibility to retain real quasi-energy of the non-unitary time-evolution operator without strong constraint on the position space. We leave this issue as a future work.

\section*{acknowledgements}
We thank Y.\ Asano, Y.\ Matsuzawa, A.\ Suzuki, and K.\ Yakubo for
helpful discussions.
This work was supported by the ``Topological Materials
Science'' (No.\ 16H00975) and Grants-in-Aid (No.\ 16K17760 and No.\ 16K05466) from
the Japan Society for Promotion of Science.

\end{document}